\documentclass[fleqn,usenatbib]{mnras}

\usepackage{newtxtext,newtxmath}
\usepackage{mathptmx}
\usepackage{txfonts}

\usepackage[T1]{fontenc}
\usepackage{ae,aecompl}
\usepackage[dvipsnames]{xcolor}

\usepackage{graphicx}	% Including figure files
\usepackage{amsmath}	% Advanced maths commands
\usepackage{amssymb}	% Extra maths symbols

\def\ah{AH~Her}
\def\m1{$M_1-M_2$}

\newcommand{\caps}[1]{{\scshape{#1}}}
\protect
\def\apj{ApJ}

\def\mnras{MNRAS}
\def\pasp{PASP}
\def\aap{A\&A}
\def\apjs{ApJS}
\def\kms{km s$^{-1}$}

\newcommand{\orcid}[1]{\textsuperscript{\href{http://orcid.org/#1}{
\hskip2pt\includegraphics[width=8pt]{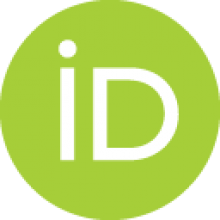}}}}

\title[Observations in quiescence of AH Her]{New radial velocity observations of AH Her: evidence for material outside the tidal radius}
\author[Echevarr\'{\i}a et al.]{J. Echevarr\'{\i}a\orcid{0000-0001-5960-3023},$^{1,2}$\thanks{e-mail: 
	jer@astro.unam.mx}
	    J.~V. Hern\'andez Santisteban\orcid{0000-0002-6733-5556},$^{3,2}$ 
    O. Segura Montero,$^{1}$
    \newauthor
    S. H. Ram\'irez,$^{1}$
    A. Ruelas-Mayorga,$^{1}$
    L. J. S\'anchez\orcid{0000-0001-9637-7774},$^{1}$   
    R. Michel\orcid{0000-0003-1263-808X},$^{4}$
        \newauthor
    R. Costero,$^{1}$ 
    D.~H. Gonz\'alez-Buitrago\orcid{0000-0002-9280-1184},$^{4}$
     and J. Olivares\orcid{0000-0003-0316-2956} $^{5}$
\\
$^{1}$ Instituto de Astronom\'ia, Universidad Nacional Aut\'onoma de M\'exico, Apartado Postal 70-264,
~Ciudad Universitaria, M\'exico D.F., C.P. 04510, M\'exico\\
$^{2}$Anton Pannekoek Institute for Astronomy, University of Amsterdam, Science Park 904, NL-1098 XH Amsterdam, the Netherlands\\
$^{3}$SUPA School of Physics \& Astronomy, University of St Andrews, North Haugh, St Andrews KY16 9SS, Scotland, UK\\
$^{4}$ Instituto de Astronom\'ia, Universidad Nacional Aut\'onoma de M\'exico, Apartado Postal 877, Ensenada, Baja California, C.P. 22830 M\'exico\\
$^{5}$Laboratoire d'Astrophysique de Bordeaux, Univ. Bordeaux, CNRS, B18N, all\'ee Geoffroy Saint-Hilaire, 33615 Pessac, France\\
}
\date{Accepted 2020 November 13. Received 2020 November 11; in original form 2020 June 29}

\pubyear{2020}

\begin{document}
\label{firstpage}
\pagerange{\pageref{firstpage}--\pageref{lastpage}
}
\maketitle

% * <jer@astro.unam.mx> 2017-05-11T12:10:41.299Z:

\begin{abstract}
Spectroscopic observations of AH Herculis during a deep quiescent state are put forward. We found the object in a rare long minima, allowing us to derive accurately the semi-amplitudes: $K_1 =121 \pm \, 4$ \kms\ and $K_2 =152 \pm 2$ \kms\ and its mass functions $M_W{ \sin }^{ 3 }i=0.30 \pm 0.01$ M$_{\odot}$ and $M_R{ \sin }^{ 3 }i=0.24 \pm 0.02$ M$_{\odot}$, while its binary separation is given by $a \sin i =1.39 \pm 0.02$~R$_{\odot}$. The orbital period $P_{orb}$~=~ 0.25812~$\pm~0.00032$~days was found from a power spectrum analysis of the radial velocities of the secondary star. These values are consistent with those determined by \citet{Horne:1986}. Our observations indicate that K5 is the most likely spectral type of the secondary. We discuss why we favour the assumption that the donor in AH Her is a slightly evolved star, in which case we find that the best solution for the inclination yields $i = 48^\circ \pm 2^\circ$. Nonetheless, should the donor be a ZAMS star, we obtain that the inclination is between  $ i = 43^\circ$ and $i = 44^\circ$. We also present Doppler tomography of H$\alpha$ and H$\beta$, and found that the emission in both lines is concentrated in a large asymmetric region at low velocities, but at an opposite position to the secondary star, outside the tidal radius and therefore at an unstable position. We also analyse the H$\alpha$ and H$\beta$ line profiles, which show a single broad peak and compare it with the previous quiescent state study which shows a double-peaked profile, providing evidence for its transient nature. 
\end{abstract}


%%%%%%%%%%%%%%%%%%%%%%%%%%%%%%%%%%%%%%%%%%%%%%%%%%

%%%%%%%%%%%%%%%%% BODY OF PAPER %%%%%%%%%%%%%%%%%%

\section{Introduction}
\label{intro}

Cataclysmic Variables (CVs) are interacting binaries, which consist of a white dwarf (WD, by definition the primary star), and a late--type secondary, a slightly evolved main sequence star filling its Roche Lobe which transfers matter to the usually more massive star via an accretion disc \citep[see][and references therein]{warner:1995}. 
Since CVs are spectroscopic binaries, the study of their radial velocities is important in order to obtain their orbital parameters e.g., mass (provided that we have a good hold on their inclination angle). Furthermore, the emission lines can be used, by means of Doppler tomography, to make a two-dimensional velocity map of the accretion disc or other components present in the emission line spectra \citep[e.g.][]{Marsh:1988}.

AH Herculis is a Z~Cam-type star. These are dwarf novae that show irregular standstills.  The unusual behaviour of Z~Cam systems is attributed to its mass transfer rate ($\dot{M}$) being close to the critical point where the accretion disc can be maintained or not as a hot and high viscosity disc. After an outburst, the enhanced mass transfer rate is slightly above the critical $\dot{M}$, which maintains a large portion of the disc ionised, producing the intermediate brightness standstills. Once the disc cools down and the mass transfer crosses below the critical threshold, the disc can become unstable  and the systems returns to its quiescent state. \citet{Buat:2001} have proposed that this disc instability  model can be explained if the mass-transfer rate from the secondary star varies about thirty percent about the value critical for stability. While this picture accounts for most of the phenomenology of Z~Cam systems, recent observations have now found outbursts occurring during long standstills \citep{Simonsen:2011,Szkody:2013}. This anomalous behaviour can be attributed also to small fluctuations in the $\dot{M}$ arising from the donor which can lead the system to undergo an outburst from its current standstill state \citep{Hameury:2014}.
Furthermore, Z-Cam-like phenomenology is not unique to white dwarf systems and might be the explanation for unusual behaviour in some black-hole X-ray binaries \citep[e.g. Swift J1753.5-0127,][]{Shaw:2019}.

In particular, during long epochs, AH~Her shows a consistent semi-periodic behaviour between quiescence and outburst, spending only one or two days at minimum, making it difficult to observe the secondary star. Given the unpredictable nature of Z-Cam systems, it is challenging to characterise them during epochs of quiescent state with only one previous radial velocity study performed on AH~Her \citep{Horne:1986}. 
In this paper, we present a new radial velocity study, based on spectroscopic observations of AH~Her taken during a deep long quiescent state. Although we also present photometry and spectroscopy taken during the previous outburst (see Section~\ref{sec:obs}), it is only due to the fact that these observations were obtained during the same run. In Section~\ref{curve-analysis} we present a synthesis of the complex light curve behaviour of this Z Cam-type object along several years. An improved ephemeris is presented in Section \ref{ephem}, while in Section~\ref{sec:radvel} we perform a radial velocity analysis for both orbital components. In Section \ref{dynamics} we discuss several dynamic parameters of AH~Her. A general discussion is made in Section~\ref{discussion}, while our conclusions are presented in Section~\ref{conclusions}.

\section{Observations}
\label{sec:obs}

\subsection{Photometry}
\label{sec:photo} 

\begin{figure*}
	\includegraphics[trim=0.2cm 0.2cm 0.0cm 0.2cm, clip,width=17cm]{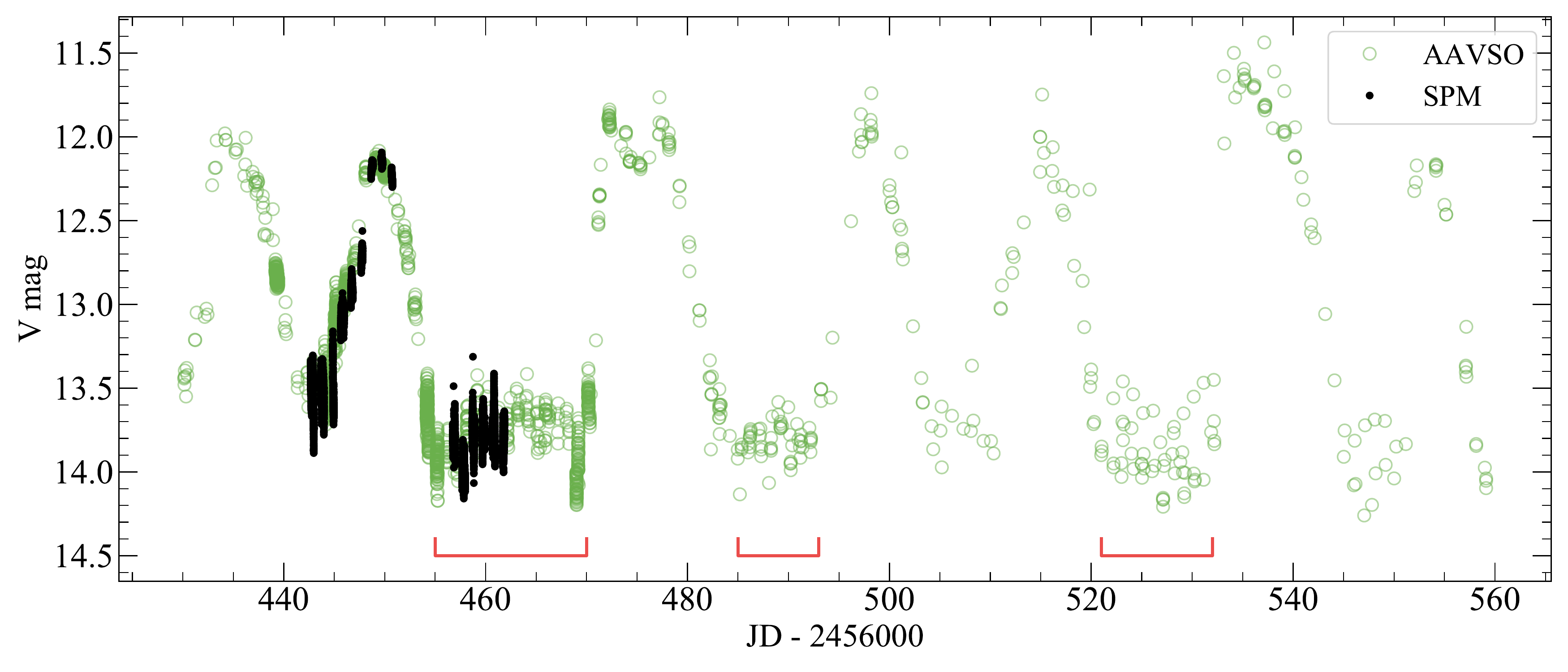}
    \caption{Optical photometry of \ah\ during high and low states taken in SPM (black) and by AAVSO (green) during 2013. Between outbursts, we observed one short and one  broad minima (the latter was used for our radial velocity analysis in quiescence presented in this work). The red bands show that this unusual broad minima was a recurrent feature during 2013.}
    \label{fig:photo}
\end{figure*}

We obtained photometric observations of AH~Her on 2013 May 30--31, June 01--07 and 14--18 at the Observatorio Astron\'omico Nacional at San Pedro M\'artir (SPM), M\'exico. We used the 84 cm telescope and the $2048\times2048$ Marconi-3 detector to perform 30 second exposures in the Johnson V band. The log of observations is presented in Table~\ref{tab:speclog}. All frames were bias and flat corrected. Afterwards, we performed differential aperture photometry using standard procedures within {\sc iraf} \citep{Tody:1986}. The photometric observations are shown in Fig.~\ref{fig:photo} as black dots. These observations are superimposed on an overall light curve, taken from the American Association of Variable Star Observers (AAVSO)\footnote{To retrieve data for AH~Her follow: \url{www.aavso.org/lcg}}, shown as  green open circles covering the time frame from day 425 to 565\footnote{JD 2456000 +}.

\begin{table}
\centering
	%\fontsize{10}{15}\selectfont 
     % se utiliza para centrar la tabla
	\caption{Log of V photometric and spectroscopic observations of AH~Her. The photometric exposure times were 30 s and 900 s for the spectroscopic observations.}
    \label{tab:speclog}
    \begin{tabular}{lccc}
        \hline
        Date & Julian Date & No of     & No of    \\
               & (2456000 +) &  Images    &  Spectra   \\
        \hline
        30  May  2013 & 442  & 726  & 28 \\
        31  May  2013 & 443  & 696  & 24 \\
        01  June 2013 & 444  & 256  & 25 \\
        02  June 2013 & 445  & 667  & 28 \\
        03  June 2013 & 446  & 358  & 17 \\
        04  June 2013 & 447  & 270  &  --  \\
        05  June 2013 & 448  & 328  &  --  \\
        06  June 2013 & 449  & 267  &  --\\
        07  June 2013 & 450  & 224  &  --  \\
        13  June 2013 & 456  & 469  &  --  \\
        14  June 2013 & 457  & 404  & 17 \\
        15  June 2013 & 458  & 420  & 23 \\
        16  June 2013 & 459  & 456  & 24 \\
        17  June 2013 & 460  & 555  & 23 \\
        18  June 2013 & 461  & 373  & 12 \\
       \hline
\end{tabular}
\end{table}

\subsection{Spectroscopy}

Spectroscopic observations were obtained with the Echelle spectrograph attached to the 2.1m Telescope of the Observatorio Astron\'omico Nacional at San Pedro M\'artir, on the nights of 2013 May 30-31, June 01--03 and 14--18. The log of spectroscopic observations is presented in Table~\ref{tab:speclog}. 
All observations were carried out with a Marconi--2 $2048\times2048$ detector and a 300~{\it l/mm} cross--disperser, which has a blaze angle around 5500~\AA, to obtain a spectral resolution around this wavelength, of R$\sim$21,000 or $\sim$14 km\,s$^{-1}$ . The final spectral coverage was about $3900$--$7300 $ \AA. The exposure time for each spectrum was 900~s. This represents a phase coverage for AH Her of $\Delta\phi\simeq0.013$. A ThAr lamp was taken every 4 exposures for accurate wavelength calibration.
All images were bias and flat corrected. The spectral wavelength calibration and extraction was performed with {\sc iraf}\footnote{IRAF is distributed by the National Optical Astronomy Observatories, which are operated by the Association of Universities for Research in Astronomy, Inc., under cooperative agreement with the National Science Foundation.}. 
%\bigskip
We observed the spectral standard stars  HD166620~(K2V), 61~Cyg~A~(K5V) and 61~Cyg~B~(K7V) for  cross-correlation purposes and rotational analysis with the donor star.

\section{An overall view at the AH~Her complex light curve}
\label{curve-analysis}

It is of general interest to look, over the years, at the overall light curve of AH~Her (a complex Z~Cam-type star) and to point out at its behaviour during our observations. We have examined the large database  of the AAVSO, particularly from 2011 to 2014, which has an excellent coverage. It would be too extensive for the length of this paper to show here the yearly light curves and discuss in detail their behaviour. However, we can describe in broad terms its main changes and characteristics, while referring the willing reader to explore the light curves at the site mentioned in Section~\ref{sec:photo}. Firstly, we note that in general, the outbursts have a maximum light of about 11.5 mag and a minimum of 14.5 mag. Secondly, we observe several combinations of sequences between maximum, standstill and minimum states, like the typical minimum--outburst--standstill--minimum sequence or a more complex standstill--outburst--standstill--minimum--outburst combination, which was observed only once in 2012 but lasted nearly four months. Thirdly, as mentioned in Section~\ref{intro}, during long epochs the behaviour of AH~Her was nearly sinusoidal during which the quiescent state lasts a couple of days only. However in 2013 we notice 3 unusually long minima as shown in Figure~\ref{fig:photo}, of which the first was partially observed by us. As mentioned in Section~\ref{intro}, we also observed a previous outburst, which had a minimum of $\sim$ 13.5 mag and a maximum of $\sim$ 12 mag. These observations, obtained from a shallow minimum up to a maximum state, although reported here (see Section~\ref{sec:obs}), are not discussed in this paper, as our purpose is to analyse the observed second deeper quiescent state in order to study the elusive secondary star. The first minimum to outburst observations will be published elsewhere.

\section{Improved Ephemeris}
\label{ephem}

Similar to the analysis of AH~Her made by \citet{Horne:1986}, we have first derived an orbital period value from a power spectrum based on our radial velocities. In our case we decided to use the secondary star results instead of the emission lines as they produce a much better result. These velocities are discussed in Section~\ref{donor}.

\begin{figure}
	\includegraphics[width=\columnwidth]{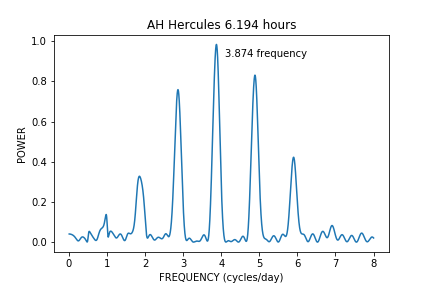}
	\caption{Power spectrum obtained by the analysis of the radial velocities of the secondary star. The maximum peak value is 3.87418 cycles/day, equivalent to an orbital period of 6.1948 hours. }
    \label{poworb}
\end{figure}

We have used a Lombe-Scargle algorithm from a {\sc python}\footnote{https://docs.astropy.org/en/stable/timeseries/lombscargle.html} package to produce a periodogram (see Fig.~\ref{poworb}). This figure shows the results obtained for our data; around the frequency value reported by \citet{Horne:1986} (3.874227 cycles/day) there appear a number of adjacent aliases. The frequency resolution we used in  the application of the Lombe-Scargle algorithm was 0.00001 cycles/day.

The best result gave a value of 3.87418~$\pm$~0.00491~cycles/day, equivalent to an orbital period of $0.25812\pm~0.00032$ days. This value is compatible, within the errors, to that found by \citet{Horne:1986}. Using this result, we fitted the data derived from Section~\ref{donor} to obtain the inferior conjunction of the secondary star to determine the new ephemeris as:
\begin{equation}
    T(HJD) = 2456457.54161 \pm 0.00051 + (0.25812 \pm 0.00032)\,E,
\end{equation}
where $HJD$ is the heliocentric Julian Day and $E$ is the phase measured from the inferior conjunction of the secondary star. 

\section{Radial velocity analysis}
\label{sec:radvel}

\subsection{The primary star}
\label{radvel1}

To measure the radial velocity of the emission lines (such that they reflect more accurately the motion of the primary star)  \citet{Schneider:1980} proposed a method to determine a more precise centre of the line by measuring the wings only; since they originate in the vicinity of the white dwarf, asymmetric-low-velocity features are avoided. This method, which uses two Gaussian functions with a fixed width and variable separation was further expanded by \citet{Shafter:1986}, who developed a diagnostic diagram. In particular, they define a control parameter, $\sigma_{K} / K$, whose minimum is a very good indicator of the best $K$ value. We have further developed this method by creating an interactive program \citep{Echevarria:2016,Hernandez:2017}, which uses a grid of width and separation to find the best values of the wings profile. At each trial combination, we fit every spectra in our sample to a circular  orbit. 
\begin{equation}
V(t) = \gamma + K_1 \sin\left(2\pi\frac{t - t_0}{P_{orb}}\right),
\label{radvel}
\end{equation}
where $\gamma$ is the systemic velocity, $K_1$ the semi-amplitude, $t_0$ the time of inferior conjunction of the donor and $P_{orb}$ is the orbital period. We employed $\chi^2$ as our goodness-of-fit parameter. Note that we have fixed the orbital period, as derived in Section~\ref{ephem}, and therefore we only fit the other three parameters. 

We have selected the spectra from nights 457--461, where the system was found at its lowest state (see discussion in Section~\ref{donor}). 
For H$\alpha$, we used a grid of widths from 150 to 300 km\,s$^{-1}$ and separations from 1040 to 1780 km\,s$^{-1}$ in steps of 20~km\,s$^{-1}$.~For H$\beta$, we used a grid for widths from 395 to 490 km\,s$^{-1}$ and separations from 1320 to 1680 km\,s$^{-1}$ also in steps of 20~km\,s$^{-1}$.

\begin{figure*}
	\includegraphics[ width=\columnwidth,trim=0cm 4cm 0cm 0cm]{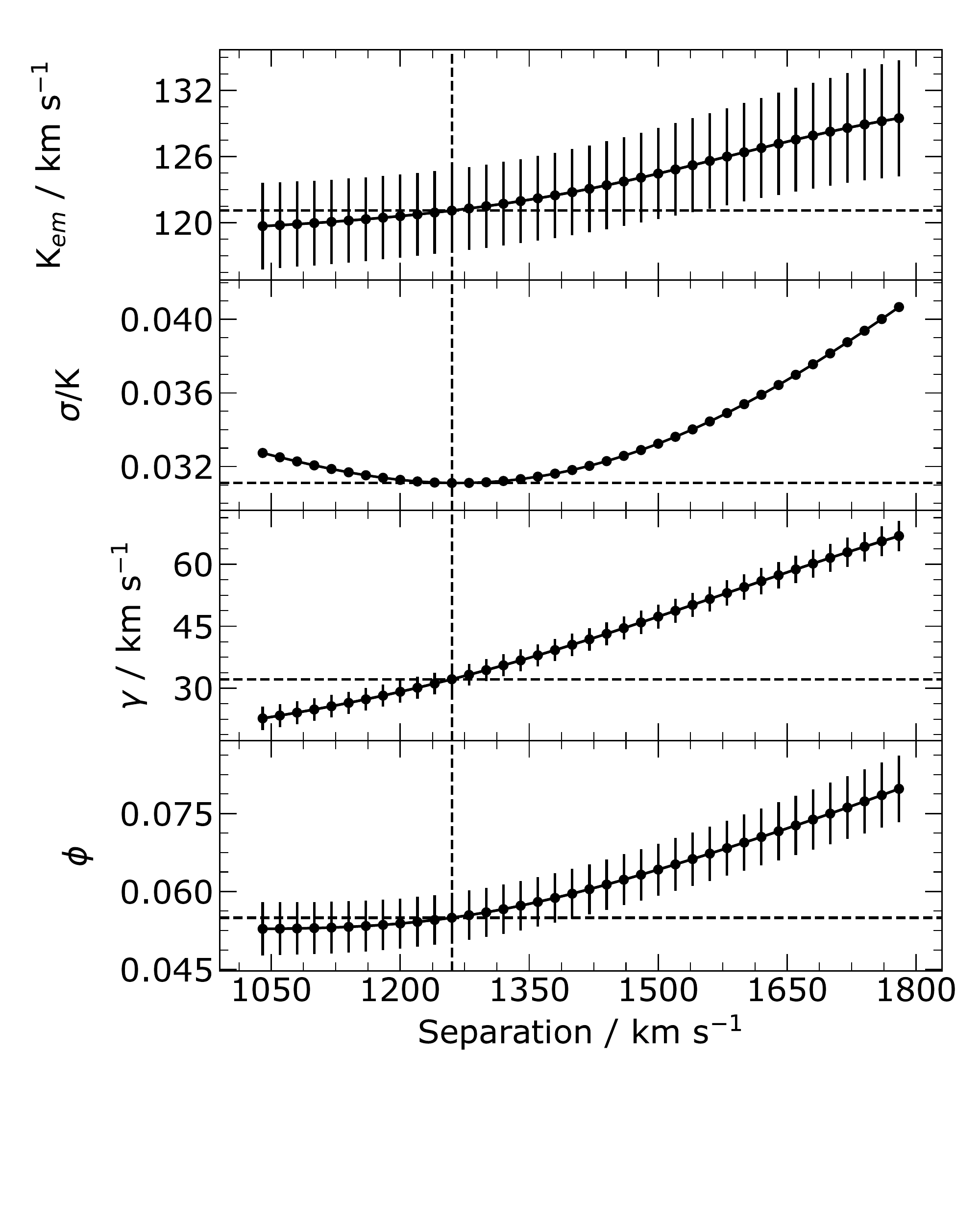}
		\includegraphics[ width=\columnwidth,trim=0cm 4cm 0cm 0cm]{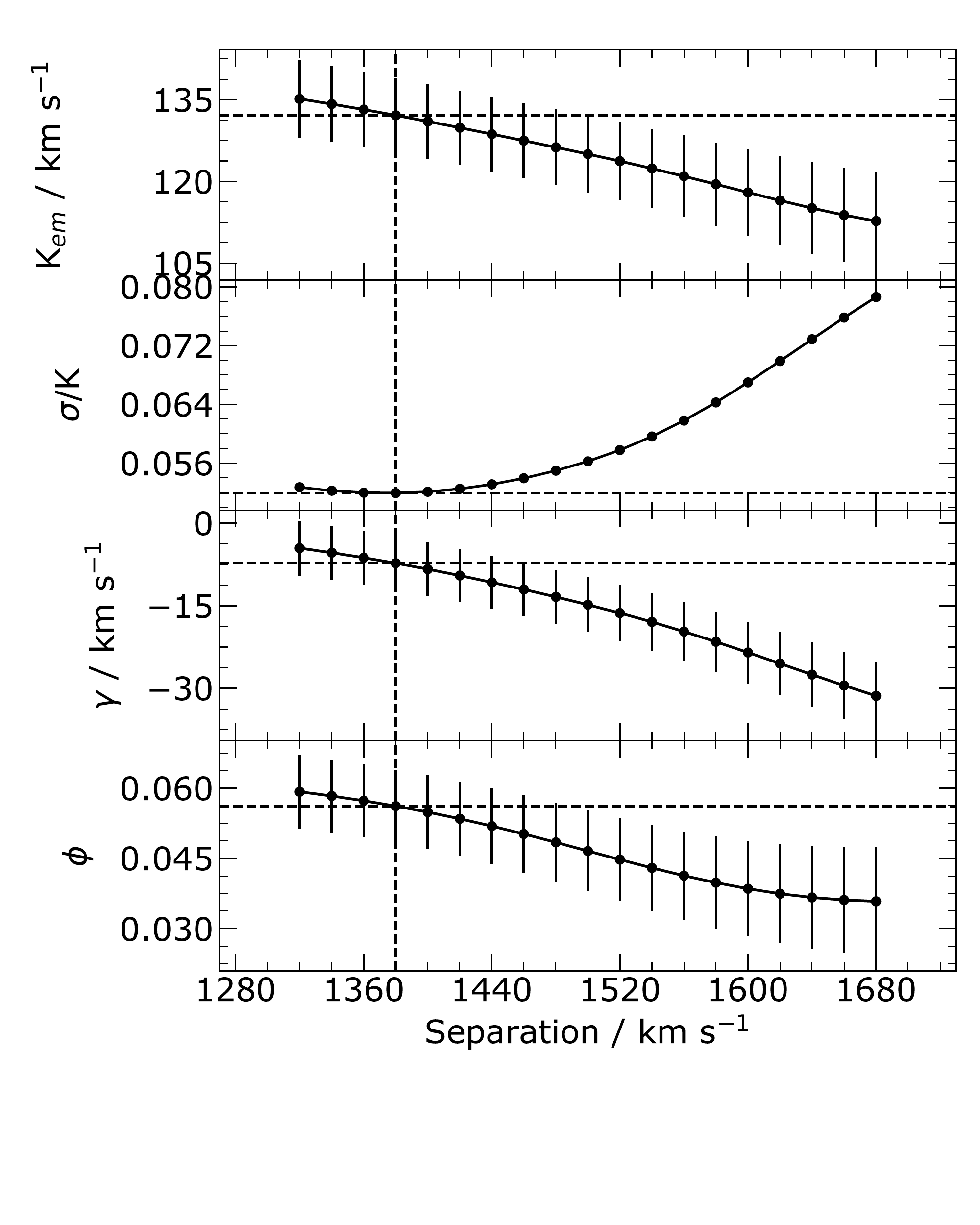}
    \caption{Diagnostic diagram for H$\alpha$ (left) and H$\beta$ (right).  The best result for H$\alpha$ is found for a Gaussian separation of 1260 \kms, which correspond to K~$\sim$ 124~\kms. The best result for H$\beta$ is found for a Gaussian separation of 1420 \kms, which correspond to K~$\sim$ 132~\kms. The orbital parameters for the solution of both lines are shown in the first and second column of Table~\ref{orbpar} respectively. The $\sigma$/K indicator shows a clear minimum in both cases. The behaviour of K, $\gamma$ and $\phi$ as functions of separation is discussed in the text.}
    \label{diag-hab}
\end{figure*}

\begin{figure}
	\includegraphics[ width=\columnwidth,trim=0.3cm 0.6cm 0cm 0.2cm,clip]{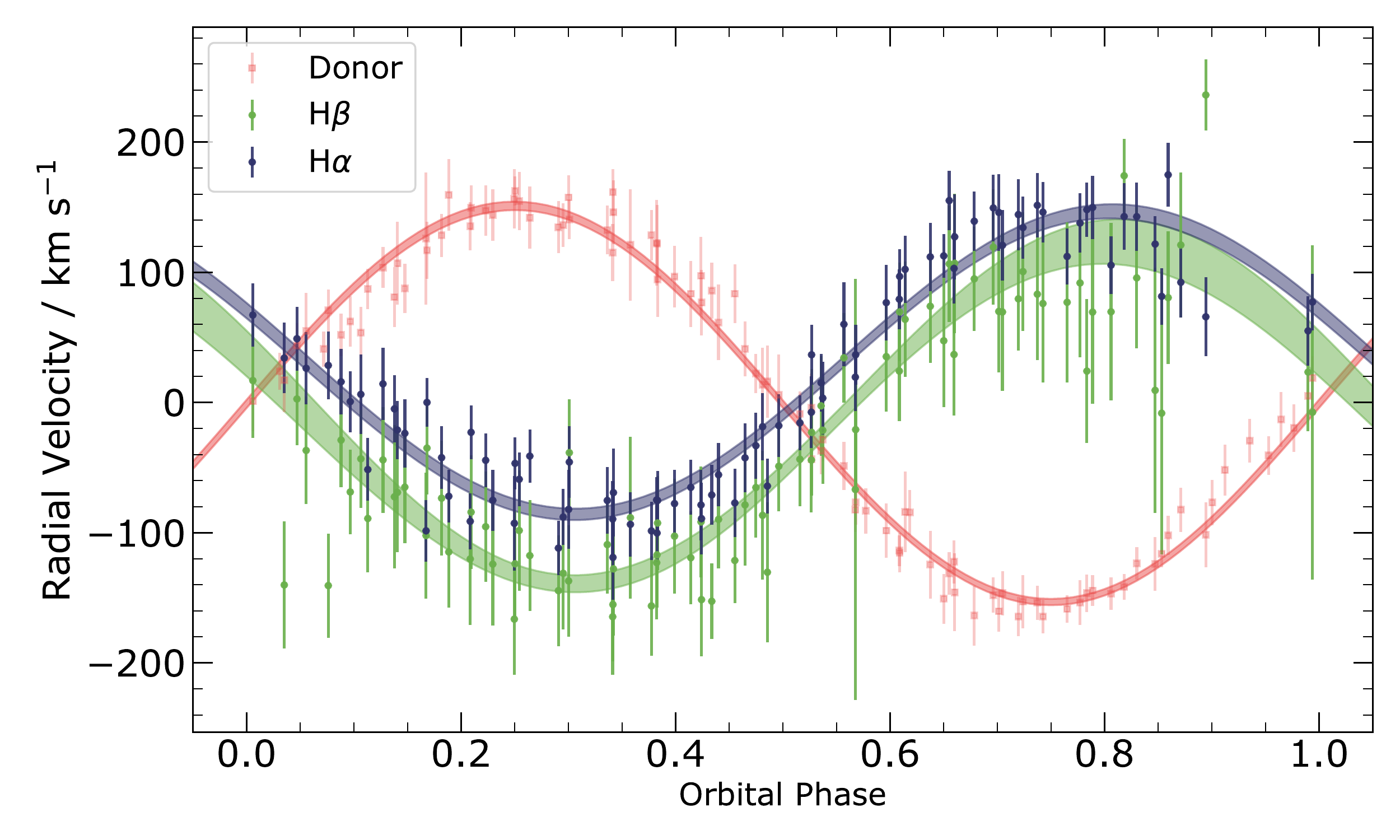}
    \caption{Radial velocity curve for the emission lines, H$\alpha$ (blue), H$\beta$ (green) and absorption lines from the secondary star (red).  The best solutions for the emission lines are shown, as well as random realisations via bootstrapping, to reflect the scatter of the solutions. Errors in individual data points have been scaled  so that $\chi^2_{\nu}=1$. The orbital parameters are shown in Table~\ref{orbpar}. Note that the crossing point between the emission and absorption solutions (i.e. at the systemic velocity) have a shift of about 0.05-0.1 in orbital phase.
    }
   \label{radvel-all}
\end{figure}

In Figure~\ref{diag-hab} we present a diagnostic diagram for H$\alpha$ (left) and H$\beta$ (right) to determine the best orbital solution by selecting the minimum of the control parameter $\sigma_{K_1} / K_1$. The optimal value is shown as a dotted vertical line. This value indicates that our best radial velocity semi amplitudes are around 120~km\,s$^{-1}$ for H$\alpha$ and 130~km\,s$^{-1}$ for H$\beta$. Figure~\ref{radvel-all} shows the radial velocity curve in a combined figure for the emission lines arising from the accretion disc and the absorption lines coming from the secondary star (the analysis and results for the donor star are discussed in Section~\ref{donor}). The blue and green lines (which includes the bootstrap region) shows the best fit models. The 1$\sigma$ error bars have been scaled so $\chi^2_{\nu}=1$. The results are also shown in the second and third column of Table~\ref{orbpar}. As mentioned by \citet{Shafter:1986}, it is instructive to plot $K_1$, $\gamma$ and $t_0$ (here shown as $\phi=t_0(a) - t_o$), as part of the diagnostic diagrams, since we expect them to behave in a particular way when approaching a good fit to the wings of the lines. These values, as well as $\sigma_{K_1} / K_1$ should show a slow approach towards the optimal values and then, as the separation increases, they should depart rapidly as the measurements are dominated by noise. i.e. we start to measure outside the wings, which is what we see in Figure~\ref{diag-hab}.

Note that H$\alpha$ and H$\beta$ give different results for the semi-amplitude. Although these values are both within the errors, we will adopt the H$\alpha$ value of $K_1=$121~$\pm$~4~km\,s$^{-1}$ as our best choice, not only because of the lower errors obtained from this line, but also because at phase 0.5, when the accretion disc is in front of us, its gamma velocity and that of the donor show the same value, near zero \kms (see Section~\ref{donor} and the analysis of the Doppler tomography in Section~\ref{sec:tomo}). Note that both emission line curves give different systemic velocities and are shifted at phase zero by about 0.05, as shown in Table~\ref{orbpar}. 
This effect has been noted by \citet{Stover:1981} in RU~Peg and in other systems, who suggested that an asymmetry in the emission distribution of the accretion discs might explain the observed features (see their Figure 4 and further discussion in our Section~\ref{sec:tomo}). 

\subsection{The donor star}
\label{donor}

Out of the ten nights of spectral observations, the first nights (442--445) were observed during  quiescence but at a magnitude brighter than the second minimum (nights 457--461). Within the period 446--456, the system entered in eruption and therefore the signature of the secondary star disappeared. Therefore, the only nights that were useful to perform a good correlation were nights 457--461. We performed a cross-correlation analysis for these nights, using orders 40 to 43, which cover the spectral region 5100 -- 5700 \AA, using our three spectral standards from K2V to K7V as templates. The best results were obtained for 61~Cyg~A (K5V) a primary spectral type star. The correlations were performed using the  {\sc fxcor} routine in {\sc iraf}. The orbital fit was performed using \caps{orbital}\footnote{Available at \url{https://github.com/Alymantara/orbital_fit}}. This is a simple least squares program that determines, in general, the four orbital parameters mentioned in  Section~\ref{radvel1}, any of which can be set to a fixed value (in our case we have set the orbital period to the value calculated in
Section~\ref{ephem}). The 1$\sigma$ error bars have been scaled so that the statistical distribution $\chi^2_{\nu}=1$. The results are shown in Figure~\ref{radvel-all} (red dots and line), while the orbital parameters are shown in the fourth column of Table~\ref{orbpar}. 

The fit to the observations is very good. The observed points shown here are the simple mean obtained from the individual results in Orders 40 to 43. The errors shown in the graph are also the root-mean-square deviation (rms) from such individual results. Our value of $K_2=152\pm2$~\kms ~is a considerable improvement from that determined by \citet{Horne:1986}, who phase-binned their spectra and obtained two very different results from their 1980 and 1981 data: $K_2 =$~148~$\pm$~10 and 175 $\pm 13$ \kms,  respectively; thus they adopted a mean value of 158~$\pm$~8~\kms (see their Figure 8 and Table 2), which agrees with our estimate within the errors. %Although our result agrees within the errors, there is no doubt that we have greatly improved the $K_2$ estimate. 
However, this improvement comes with a downside, since obtaining a radial velocity curve and semi-amplitude value with higher precision, reveals a new problem. We can see this in Figure~\ref{radvel-all}, where several points (red) show small deviations along some orbital phases of the calculated circular orbit. These deviations could indicate that the secondary star has an ellipsoidal shape rather than a spherical one. This is well expected in cataclysmic stars \citep[e.g.][]{Hellier:2001}. Moreover the late type star could have hot or cold spots that would affect the radial velocity curve. These effects have clearly been detected in the case of AE~Aqr \citep{Echevarria:2008,Hill:2016}. The first authors made a detailed study of the ellipsoidal variations as a function of orbital phase; while the latter found clear evidence of spots and the presence of sinusoidal residuals in the radial velocity curve. In the case of AH~Her we have a veiled absorption spectrum and we are unable to perform similar analyses as those made in AE~Aqr. However, we should caution the reader that the $K_2$ value and error would be affected by these problems and therefore  the estimated error of 1.5 percent could well be underestimated.

We should also consider and discuss whether the donor star is a ZAMS star or slightly evolved, as this matter will have implications in our calculation of its mass (See Section~\ref{sec:diag}). Early in the eighties \citet{Echevarria:1983} found that secondaries in CVs are, in general, later spectral types rather than main sequence stars of the same mass. They pointed out that {\it by studying their spectral types one may show that the secondaries are not normal main sequence stars, certainly not so for the longer periods}. \citet{Beuer:1998} refined this result, finding that CVs {\it with orbital periods less than 3~hr are close to solar abundance stars as defined by single field stars, while for periods greater than 3~hr, the earliest spectral types at a given period correspond to main sequence stars, while the majority of secondaries have later spectral types}. These authors also calculate models for evolved sequences and compared them with evolutionary ZAMS models (see references within their paper). Their conclusion is that {\it an unexpectedly large fraction of CVs has an evolved donor, and that mass transfer rates for
periods greater than 4~hr could be much higher than usually assumed}. In a more recent study, \citet{Knigge:2006} derived an empirical spectral type--orbital period relation for main sequence stars and CVs (see their figure~7) where clearly the spectral types of CVs with periods greater than 4~hr scatter to later types than those obtained for main sequence stars, and as the orbital period increases, so does the extent of such deviations; the same occurs when comparing the CVs with a main sequence theoretical relation arising from a 5-Gyr isochrone obtained by \citet{Baraffe:1998}, plotted in the same Figure. \par

Furthermore, \citet{Kolb:2000}, aside from deriving the mass-spectral type relations (discussed in Section~\ref{sec:diag}),  also show a spectral type--orbital period relation, where they compare the data taken from \citet{Beuer:1998} with their own ZAMS sequence, where again, for periods larger than 6~hr, the CVs depart more from the main sequence as the orbital period increases. Due to the results discussed above we are inclined to favour a somewhat evolved secondary star in AH~Her, as this system has a period greater than 6~hr.

\begin{table}
\centering
\caption{Orbital Parameters for AH~Her for the primary and secondary stars. The calculations were performed with a fixed orbital period of 0.25812 days.}
\label{orbpar}
\resizebox{0.45\textwidth}{!} {
\begin{tabular}{llll}
\hline
Parameter  &  H$\alpha$     &     H$\beta$ & Donor                \\
\hline
$\gamma$ $(km\,s^{-1}$) & 32      $\pm$ 3          & -7     $\pm$ 7         & -1   $\pm$ 2     \\
K ($km\,s^{-1}$)      & 121     $\pm$ 4          & 132     $\pm$ 11         & 152  $\pm$ 2   \\
$\Delta$ $\phi$            & -0.054   $\pm$ 0.005       & -0.05   $\pm$ 0.01      & ---                  \\
\hline
\end{tabular}
}

\end{table}

\section{Dynamics of AH Herculis}
\label{dynamics}

\subsection{Rotational broadening, mass ratio and inclination angle}
\label{broadening}
It is well established that in CVs the tidal force on the secondary star not only eliminates any eccentricity of the orbit, but also causes the star to co-rotate with the orbital period \citep[e.g.][and references therein]{warner:1995}, i.e. the translational angular velocity and the rotational velocity are equal: $\omega_2(trans)=\omega_2(rot)$.Thus, it is relatively easy to derive the following equation:
\begin{equation}
{ V }_{\textrm{rot}}\sin i=K_R\,(1 + q)\,\frac{{R}_{R}}{a}, 
\label{rot-bas}
\end{equation}
where $K_R$ is the radial velocity semi-amplitudes of the  secondary star and $R_R/a$ is the the mean Roche radius in units of the binary separation. Given the derived values for the semi-amplitudes, we have that $q = 0.80 \pm 0.03$ and therefore,
from Eq.~\ref{rot-bas} we obtain:

\begin{equation}
{ V }_{\textrm{rot}}\sin i=99 \pm 2 ~\textrm{km\,s}^{-1}, 
\label{rot-bas2}
\end{equation}

where we have used the value of $R_R/a$ = 0.36 $\pm$ 0.01, derived from \citet{Echevarria:1983}. Using  \citet{Eggleton:1983}, yields the same result $R_R/a$ = 0.36 $\pm$ 0.01.  The above errors, and all subsequent propagation of errors in the paper have been calculated using a standard Monte Carlo simulation.

\begin{figure}
	\includegraphics[trim=1.6cm 0.6cm 0.4cm 0.3cm, clip,width=\columnwidth]{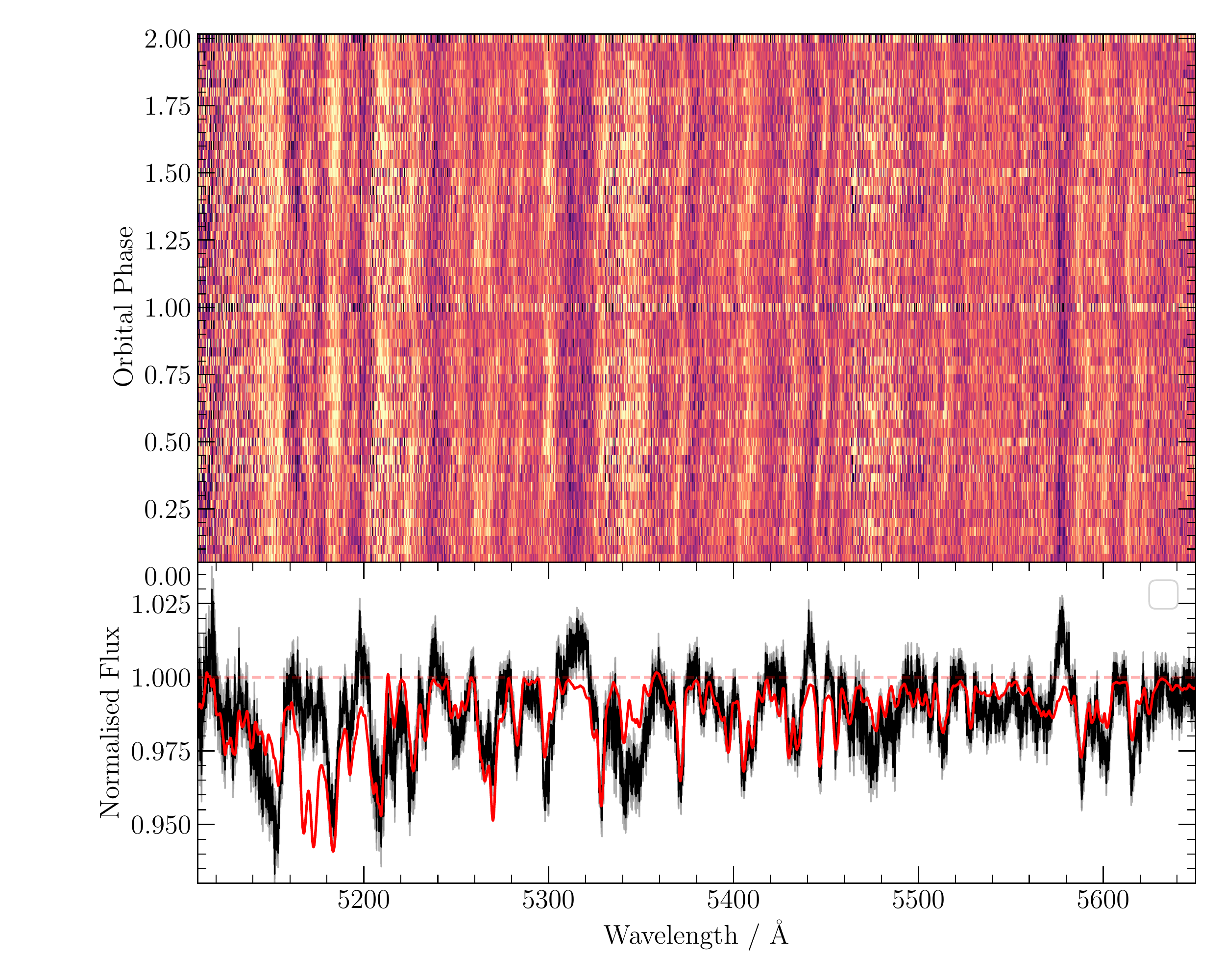}
    \caption{Detection of the donor star in AH Her. \textit{Top:} The phase-folded trail spectra show clear sinusoidal modulation of absorption lines arising in the donor. \textit{Bottom:} The continuum-normalised average spectrum shifted to the reference frame of the donor (black) is fitted by a broadened K5V template (red). }
    \label{fig:vsini}
\end{figure}

We can also obtain an independent value of the rotational velocity $V_{\textrm{rot}}\sin i$ by measuring the broadening of the absorption lines of the secondary. This can be done if we compare the lines with appropriate broadened templates of standard stars \citep[e.g.][]{Echevarria:2008}. To obtain these calibrated broadened templates, we first  proceeded to normalise and combine four individual Echelle orders of the spectrum (orders 40 to 43), which cover the wavelength region $\lambda\lambda$ 5100--5700 \AA. The expected sinusoidal behaviour of the absorption lines as a function of orbital phase is shown in the top panel of Fig.~\ref{fig:vsini}. Next, we used the solution obtained in Section~\ref{ephem} to correct every spectrum and produce a high signal-to-noise average spectrum in the reference frame of the donor, shown in the bottom panel of Fig.~\ref{fig:vsini}. As mentioned before, since the donor is tidally-locked, its rotation will broaden the absorption lines significantly. Making use of spectral templates obtained with the same instrumental setup, we performed a non-linear fit to the data by convolving the template spectrum with a rotational velocity kernel \citep{Gray:2005} and optimising $v\sin i$ as well as the fractional contribution of the donor to the overall light. This latter parameter dilutes the amplitude of the lines since the the donor only contributes a fraction of all the light at those wavelengths. We performed this fit for our K2V, K5V and K7V templates, all of which have very small intrinsic rotations compared to the donor star. The results are shown in the lower panel of Figure~\ref{fig:vsini}. 

We found that the best value is for the K5V star, although with high values of $\chi^2_{\nu} = 4.92$ (opposed to 5.04 and 5.05 for the K2V and K7V, respectively). These high values in $\chi^2_{\nu}$ are mainly driven by artificially created jumps in the normalisation of the template spectra seen as large residuals in all three templates. These occur at those points where the orders overlap, and are caused by the blue part of the combined orders having a small signal due to the blaze behaviour of the spectrograph, therefore distorting the normalised co-added spectrum. Because of the above arguments we will formally adopt a K5 spectral type, but will also take a glance at the adjacent types K4--K6, to explore the mass range of the secondary in Section~\ref{sec:diag}.

Subtracting the square values of the rotation of AH Her and the Echelle's instrumental profile ($\sim 14$ \kms), we obtain $v\sin i =~103~\pm~3$~\kms and a fractional contribution of the donor to the total light of $9.2 \pm 0.1 $\%. We note that the 1$\sigma$ uncertainties associated to these measurements do not include any systematic effects of the spectral type, so we expect our real uncertainties to be larger. Despite this fact, we measured a $v\sin i$ value that is consistent within the uncertainties to the rotational value found above, i.e that is compatible with our derived mass ratio $q = 0.80 \pm 0.03$.

\subsection[]{The $M_1-M_2$ diagram}
\label{sec:diag}

\begin{figure}
	\includegraphics[trim=0cm 1cm 0cm 2cm,clip,width=\columnwidth]{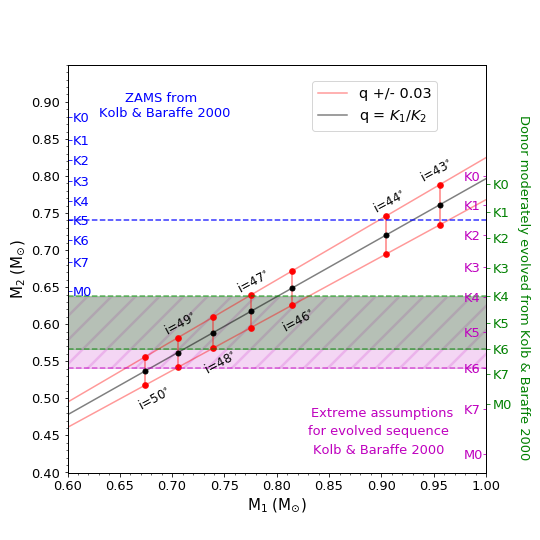}
    \caption{$M_1-M_2$ mass diagram for AH Her. The slopes of the lines are given by $q=K_1/K_2$; the black slope is for q = 0.80 and the red slopes represent its error ( $\pm 0.03$). The spectral types of the secondary star on the vertical right-hand side, correspond to moderately evolved main sequence stars (green), while those with extreme assumptions for evolved main sequence stars are shown in violet \citep{Kolb:2000}. In  the left-hand side we show (blue) the ZAMS sequence also taken from  \citep{Kolb:2000}. The horizontal green highlighted area indicates the range K4--K6, for the moderately evolved sequence; while the violet area with diagonal hatch lines, corresponds to the same spectral range but for donor stars with extreme assumptions for an evolved sequence. The connected red-black-red  dots correspond to the relevant inclination angles within these areas. We have also included a horizontal blue dashed line corresponding to a K5 ZAMS star.}
    \label{fig:m1-m2}
\end{figure}

A tool for the analysis of the masses and the angle of inclination was developed by \citet{Echevarria:2007}. This diagnostic diagram $M_1-M_2$ is shown in Figure \ref{fig:m1-m2}. In this diagram, the slopes of the lines are given by $q$ (black) and the calculated errors (red); therefore at any point in this line, for a given value of $M_1$ there is a corresponding value of $M_2$. The spectral types corresponding to moderately evolved stars that are still within the main sequence, are shown in green  on the right-hand side of the $M_{2}$ axis; while those with extreme assumptions for evolved main sequence stars are shown in violet \citep[taken from][]{Kolb:2000}. On the left-hand side, a ZAMS sequence obtained from the same authors, is shown in blue. The horizontal highlighted areas correspond to the range K4--K6 (see previous section) for each of the two different evolved models. We also included a horizontal blue dashed line to highlight a K5 ZAMS (see below). The connected dots correspond to relevant inclination angle values for the aforementioned slopes of $q = 0.80$  and for the errors $\pm 0.03$.  Within this restrictions and relationships, this $M_1-M_2$ diagram is a useful tool for the analysis of the masses and inclination angle in CVs, provided we have information on $q$ and on the spectral type of the secondary. 

Given our discussion of the rotational broadening in the secondary star (Section~\ref{broadening}), we arrived at the conclusion that the best spectral type of the secondary corresponds to a K5 star. If the secondary star is a ZAMS star then, following the $M_1-M_2$ diagram, the mass of the secondary should be around 0.74~$M_{\odot}$. In this case the inclination angle of the system would be $\sim44^{\circ}$, and the mass of the primary would be about 0.93 M$_{\odot}$. However, as there is plenty of observational and theoretical evidence (see Section~\ref{donor}) that the secondary could have moved above the ZAMS sequence, we examine the consequences of having a moderately or even highly evolved main sequence star using the models of \citet{Kolb:2000}. We explore now not only the results for a K5 star, but also for the nearby classifications K4--K6. These spectral ranges set the limits shown as green and violet bands in the $M_1-M_2$ diagram (Fig.~\ref{fig:m1-m2}). The inclination angles that correspond to these regions, vary from 46 to 50 degrees with a probable mean value at 48$^\circ$. We believe it is safe to assume $48^\circ \pm 2^\circ$. In this case we obtain a secondary mass of about $0.59 \pm 0.05~M_{\odot}$ and a primary mass of about $0.74 \pm 0.06~M_{\odot}$. Note that the difference for a K5 evolved star either moderately or extreme will only make a difference of $0.02 M_{\odot}$. In view of the arguments put forward in Section~\ref{donor} we are inclined to favour these latter results for the evolved models. Further discussion on the masses and orbital separation will be given in Section~\ref{orbparam}.

\subsection{Dynamical masses and orbital separation}
\label{orbparam}

To obtain a measurement of the masses for the AH Her binary, the emission and absorption line radial velocities are interpreted as measurements of the orbital motions of the primary and the secondary stars (white dwarf and red dwarf). Therefore, the most important parameter for this study are the semi-amplitudes of the radial velocity curves. Since this is not an eclipsing system, only the mass functions and $a \sin i$ can be determined. The inclination angle and the masses can only be estimated making additional assumptions as discussed in Section~\ref{sec:diag}.

Our measurements of AH Her were made during a deep quiescent stage. We thus assume no heating effects from the primary or disc affect the donor star and also that there are no hot or cold spots present in the secondary. Likewise, we assume here that the wings of the Hydrogen lines are not affected by any asymmetries. Then we can assume that $K_1 = K_w$ and $K_2 = K_R$. Thus, if we  adopt $K_{w}= 121 \pm 4$ \kms, $K_{R}= 152 \pm 2$ \kms, $q = 0.80~\pm~0.03$ ~and   $P_{orb}= 0.2581 \pm ~0.0003$ d, then the mass functions become:

\begin{equation}
    { M }_{ w }{ \sin }^{ 3 }i=\frac { P{ K }_{ R }({ { K }_{ w }+{ K }_{ R }) }^{ 2 } }{ 2\pi G }=0.30 \pm 0.01 M_{\odot}
\end{equation}

\begin{equation}
    { M }_{ R }{ \sin }^{ 3 }i=\frac { P{ K }_{ w }({ { K }_{ w }+{ K }_{ R }) }^{ 2 } }{ 2\pi G }=0.24 \pm 0.02  M_{\odot},
\end{equation}

\smallskip

while the separation between the two stars is given by:

\begin{equation}
    a \sin i =\frac { P({ K }_{ w }+{ K }_{ R }) }{ 2\pi  }=1.39 \pm 0.02 R_{\odot}
\end{equation}

\begin{figure*}
	
	\includegraphics[height=6cm,width=1.0\columnwidth]{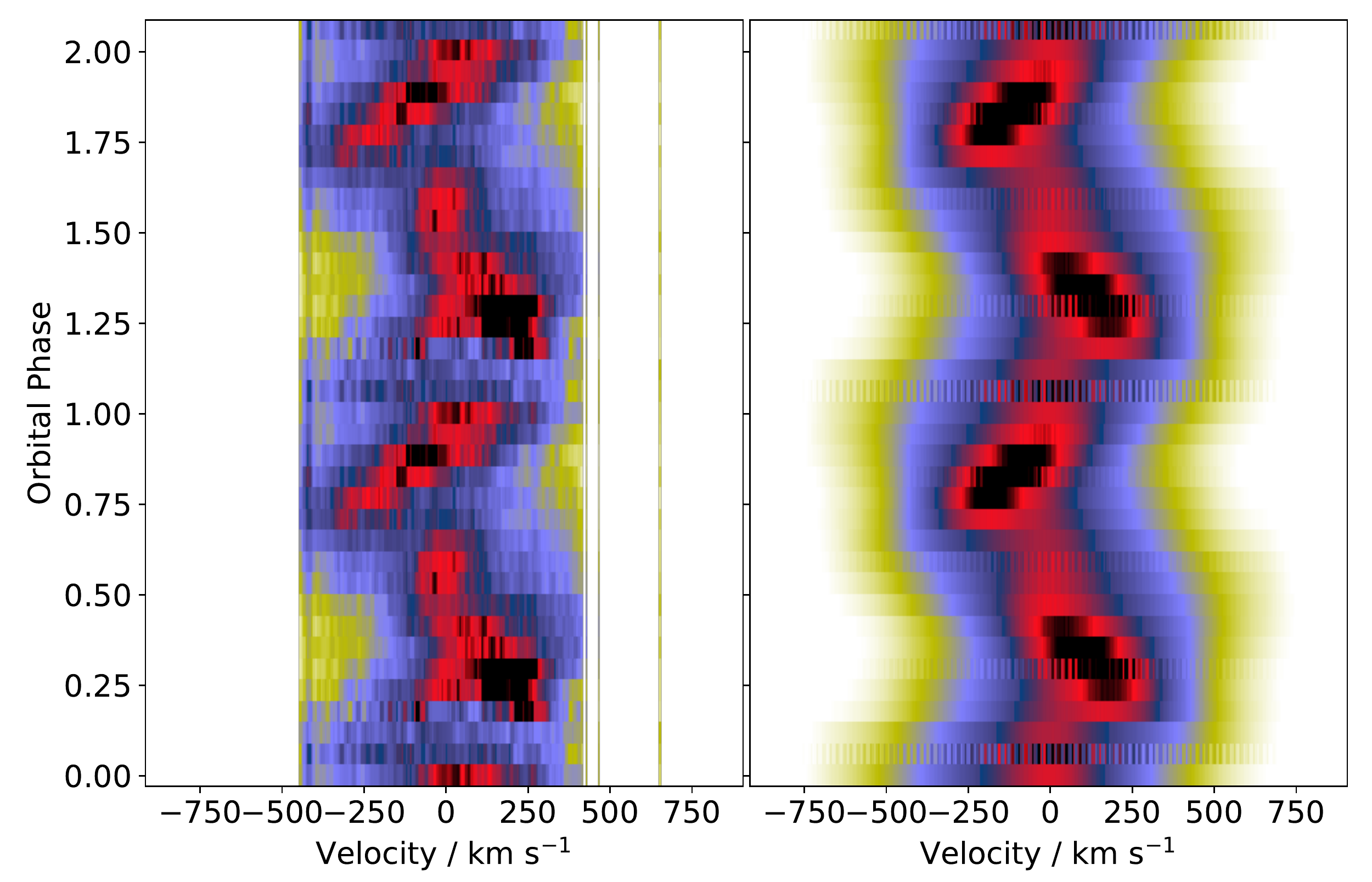}
	\includegraphics[width=1.0\columnwidth]{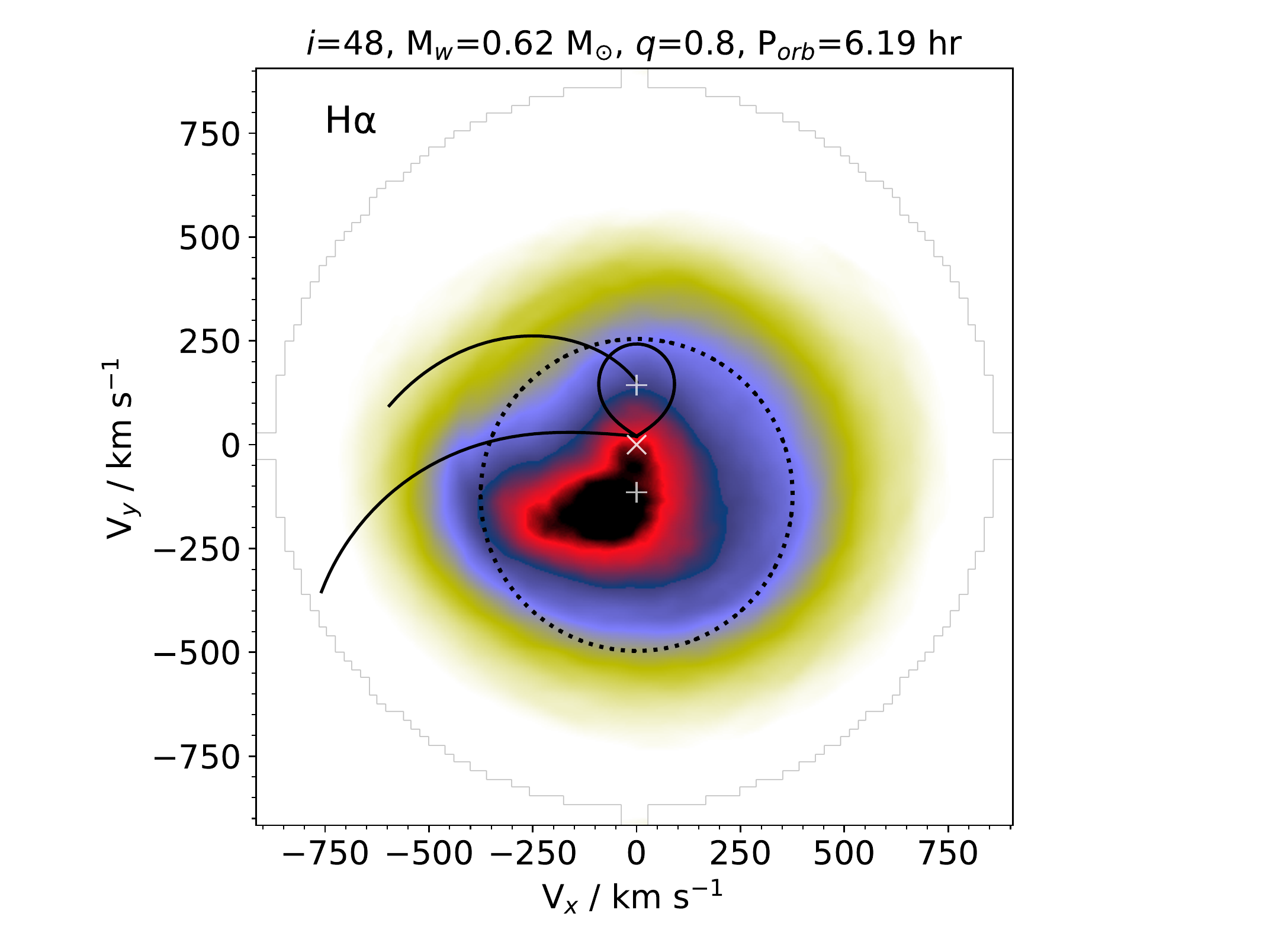}
	
	\includegraphics[angle=0,height=6cm,width=1.0\columnwidth]{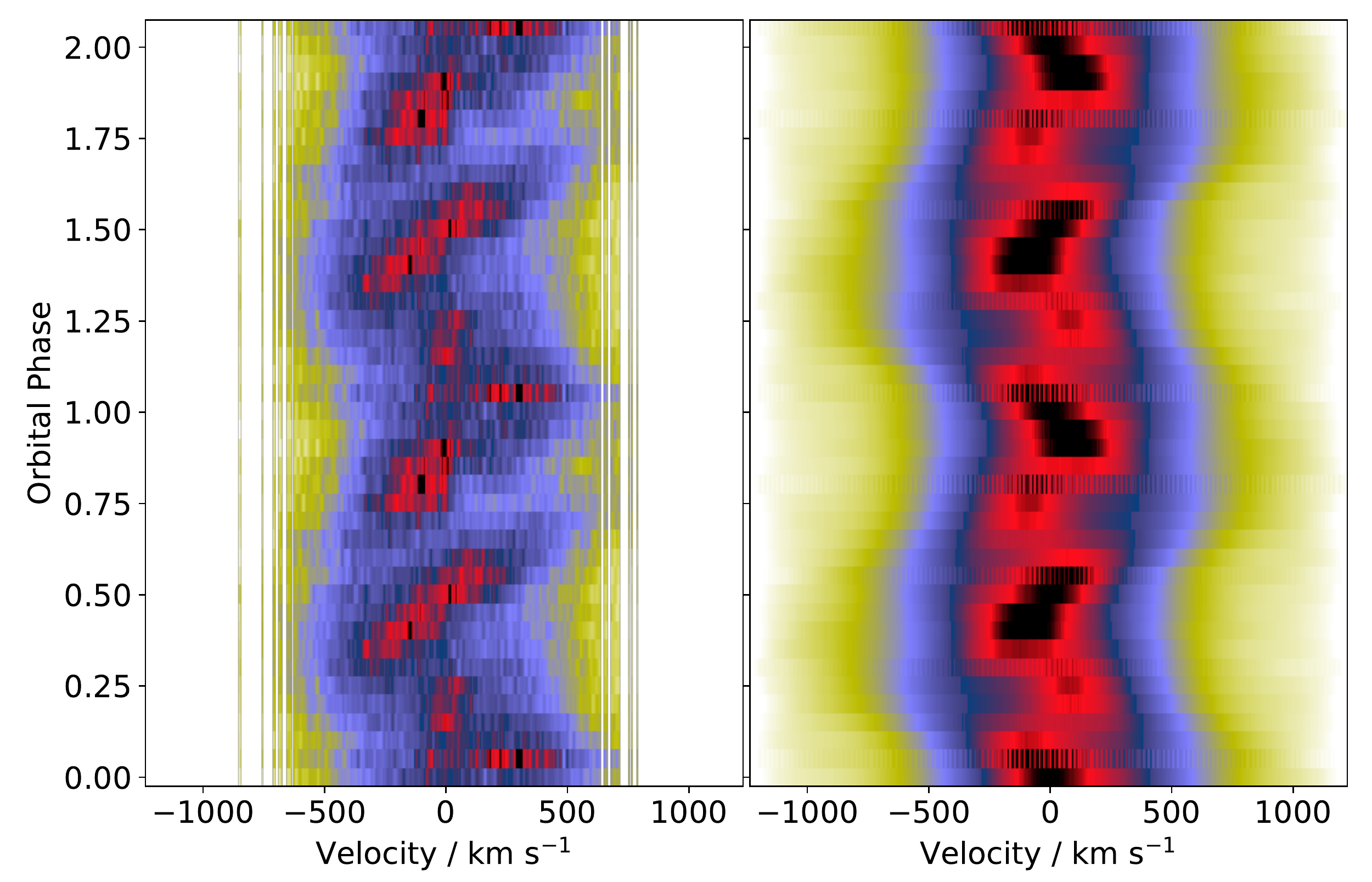}
	\includegraphics[angle=0,width=1.0\columnwidth]{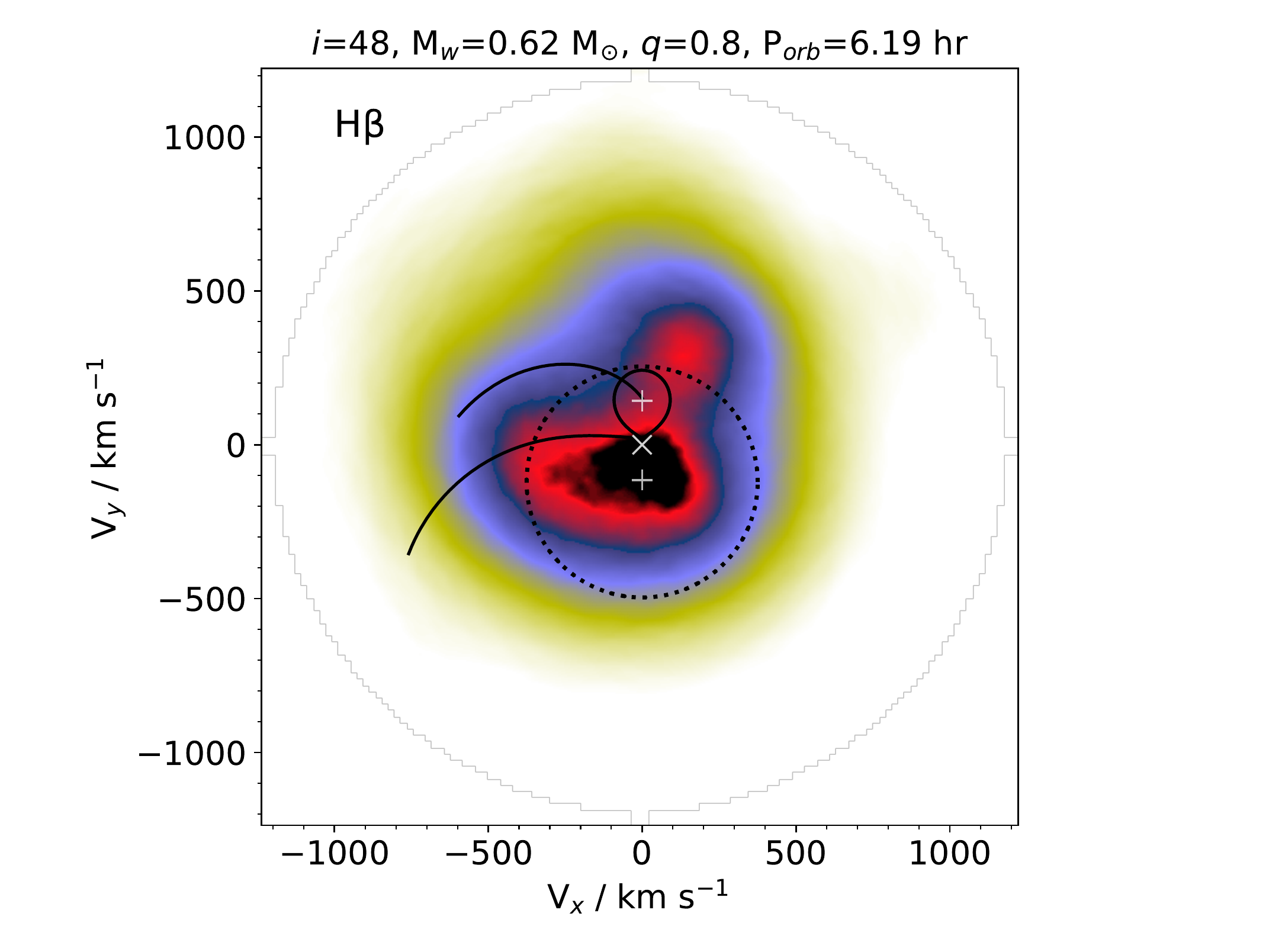}
    \caption{\textit{Top:} Trailed spectrum of the H$\alpha$ emission line for the observed data ({\em left}), the reconstructed trailed spectrum ({\em middle}), and the Doppler Tomography ({\em right}). 
    \textit{Bottom:} Trailed spectrum of the H$\beta$ emission line for the observed data ({\em left}), the reconstructed trailed spectrum ({\em middle}), and the Doppler Tomography ({\em right}). The relative flux is depicted in a scale of colours, where black represents the highest intensity, followed by red, then blue, and finally yellow. Various features in the tomographies are marked as follows:  the crosses are the velocities (from top to bottom) of the
secondary star, the centre of mass and the primary star. The Roche lobe of the secondary is shown around its cross. The Keplerian and ballistic trajectories of the gas stream are marked as the upper and lower curves, respectively. The dotted circle defines the limit given by the tidal radius. }
   
    \label{dopmap-hab}
\end{figure*}

\section{Discussion}
\label{discussion}

\subsection{Doppler Tomography}
\label{sec:tomo}

A well known time-resolved spectroscopy technique, Doppler tomography, is based on the use of a detailed emission line profile, in order to map the accretion flow in velocity space. A detailed formulation of this technique can be found in \citet{Marsh:1988}. We used the H$\alpha$ and H$\beta$ lines, to produce Doppler tomograms using a newly developed wrapper \caps{pydoppler}\footnote{Available at \url{https://github.com/Alymantara/pydoppler}}. This code uses the {\sc fortran} programs, originally developed by \citet{Spruit:1998}\footnote{Available at \url{https://wwwmpa.mpa-garching.mpg.de/~henk}} for an {\sc idl} environment. 

In the upper left panels of Figure~\ref{dopmap-hab}, we show the observed trailed spectrum ({\sc spec}) and the reconstructed trailed spectrum ({\sc reco}) of the H$\alpha$ emission line profile, while the Doppler Tomogram ({\sc dopmap})  is displayed in the upper right panel. With the same layout, we show in the lower panels of the same Figure, the Doppler tomography results based on the profile of the H$\beta$ emission line.

The observed trailed spectrum of H$\alpha$ exhibits a broad sine wave behaviour, where the highest intensities of emission are concentrated in the regions of lower velocities. The intensity decreases for higher velocity values. The spectra present a single-peaked structure, and the velocities are consistent with those observed for $H\alpha$ in Figure~\ref{radvel-all}. Note that at phase 0.5 the maximum intensity has near zero velocity, similar to that of the donor star, which is part of the reason (argued in Section~\ref{radvel1}) to adopt the semi-amplitude results from H$\alpha$. The reconstructed spectrum reproduces the main features of the observed trailed spectrum rather well. The Doppler tomography is dominated by a blob centred in  the lower-left quadrant at $(V_x,V_y)\sim(-50,-150)$ \kms.  \par

The trailed spectrum of H$\beta$ shows a different behaviour. There is a broader component (blue) which has a sinusoidal behaviour showing a minimum velocity at phase 0.25 and a maximum at phase 0.75. There is, however an extra positive velocity feature at phase 0.0. The central and brightest feature (red) does not show a sinusoidal behaviour. It rather goes from negative to positive velocities (from phases 0.25 to 0.75) and then abruptly disappears. The strong feature reappears briefly at phase 0.8, and then at phase~0.0 as a blurred feature. Finally it also shows around phase 0.15 with zero velocity.  The reconstructed spectrum shows unrealistic features, particularly a sinusoidal section around phase 0.0. This, we believe is the result of unreal features produced in the tomogram. In fact, H$\beta$ is an emission line with an erratic behaviour. Although most of the individual spectra appear single-peaked, there are a significant number that exhibit a double-peaked structure. There is also, at times an intense s-wave signal that oscillates in anti-phase with the rest of the profile. This could be the feature seen at phase 0.8, 0.0 and 0.15. Such anti-phase s-wave is the likely cause that leads the H$\beta$ tomography  to display a circular region of high intensity centred at $(V_x,V_y)\sim(+125,+300)$ \kms.
This tomogram, like that of H$\alpha$, also exhibits an asymmetric blob-like region in the bottom quadrants that reaches a local maximum at $(V_x,V_y)~\sim~(50,-100)$.

The asymmetric blobs in the low velocity region of the bottom quadrants of both tomograms might be the cause of the phase shift and systemic velocities discussed in Section \ref{radvel1} as proposed by \citet{Stover:1981}, who in fact depict a similar blob-like feature in the bottom quadrant (see his Figure 4). Moreover, this  blobs resemble those generated by SW Sex type stars \citep[e.g.][]{Schmidt:2017, Rodriguez:2007}, which are thought to be caused by material overshot by the ballistic stream, that lands on the far-side of the disc, creating a secondary impact point. Now, given the long quiescent state in which we observed AH Her, it is unlikely that the mass-transfer rate is comparable to that in  SW Sex stars; but since it is close to the transition between a transient and a steady state disc, it is possible that the stream can overflow in a similar fashion to those systems. 

Another possibility is emission from outside the orbital plane, such as an outflow. Material being ejected from the system in ballistic trajectories can collide far out of the binary plane. Such collisions can create emission regions with slower velocities often observed in the third and fourth quadrant of the Doppler tomography, as observed in AE~Aqr \citep{Eracleous1996,Wynn1997}. 

\subsection{Material outside the tidal radius?}

\begin{figure}
	\includegraphics[ width=\columnwidth]{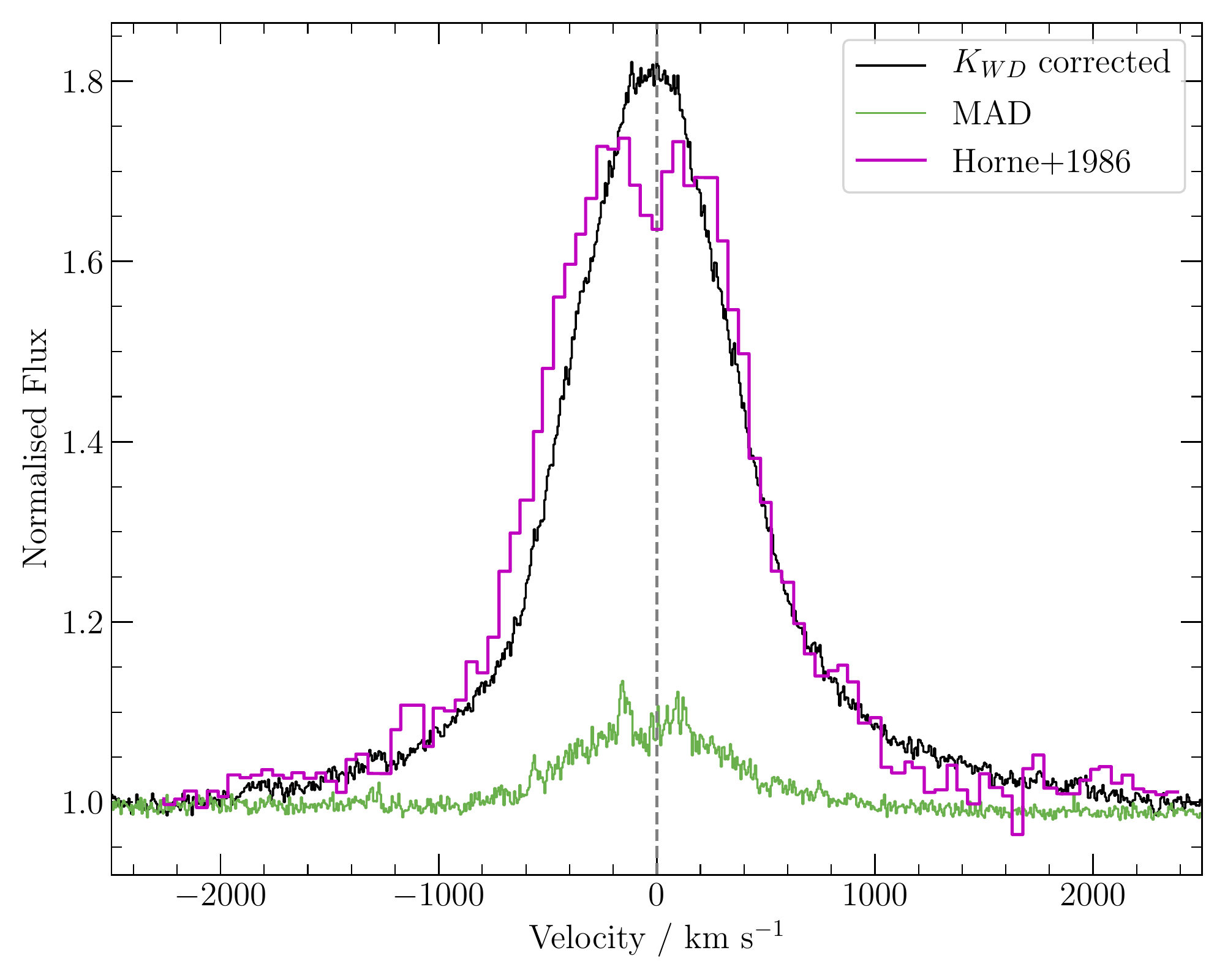}
	\includegraphics[ width=\columnwidth]{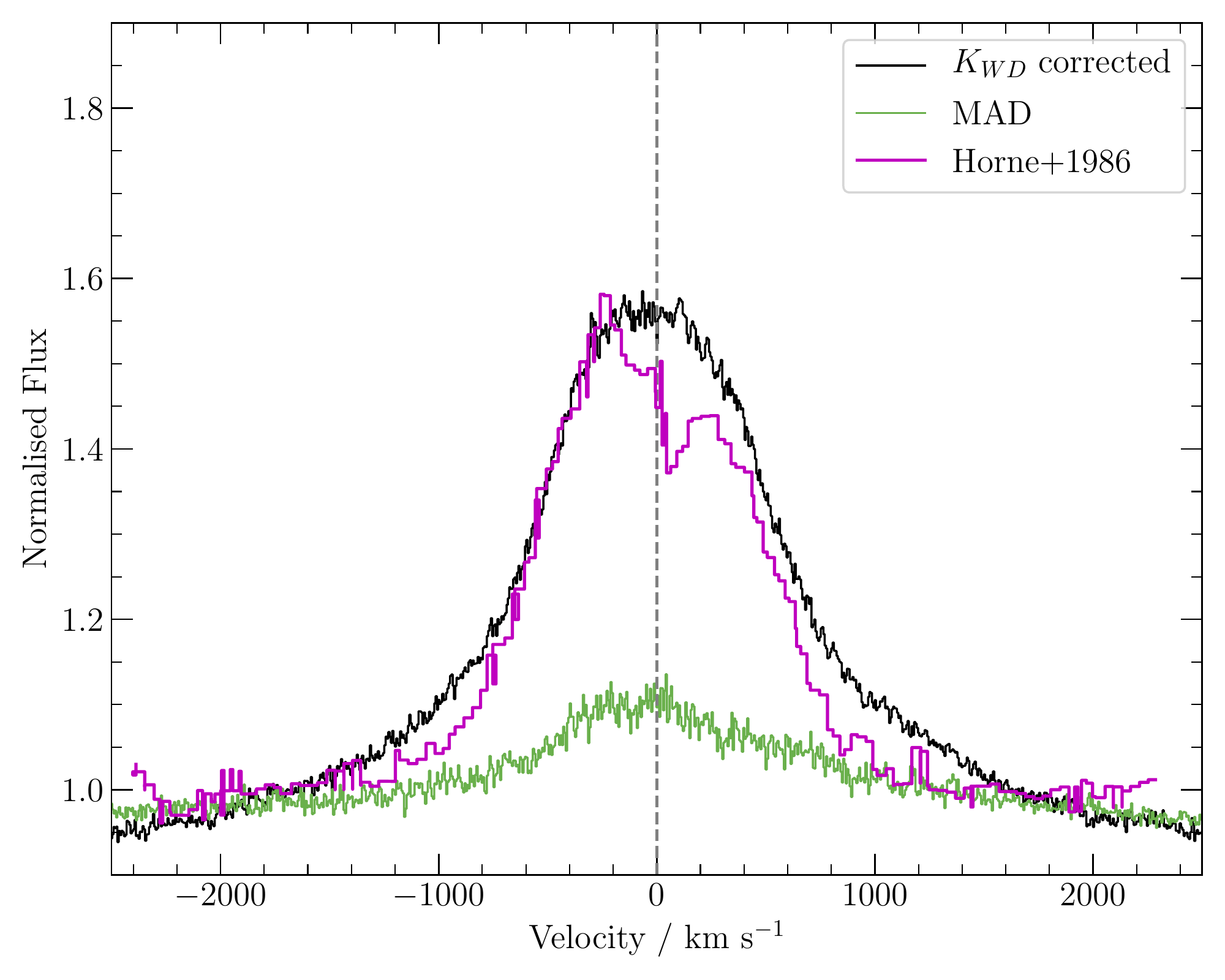}
    \caption{H$\alpha$ ({\em top panel}) and H$\beta$ ({\em bottom panel}) line profiles in the reference frame of the WD, using the ephemeris obtained in Sec.~\ref{sec:radvel}. We show the median value (black) and the median absolute deviation (MAD, shown in green). The MAD reflects the variability of the line observed in 2013 which mostly appears as a single peak component. As a comparison, we reproduce the median profile obtained in a previous quiescent state (purple) by \citet{Horne:1986} which shows a clear double-peaked emission.}
    \label{line_profile}
\end{figure}

We notice a substantial difference in the line profile between our observations and those performed by \citet{Horne:1986}. In particular, our spectra do not exhibit a strong double-peaked morphology. We explicitly show this comparison for H$\alpha$ (upper panel) and for H$\beta$ (lower panel) in Figure~\ref{line_profile}. The wings of the line are remarkably similar in H$\alpha$  between these two epochs, however, we note that the main difference arises in the core. We can separate this core component by constructing a median spectrum in the reference frame of the WD, using the orbital solution obtained in Section~\ref{sec:radvel}. We calculated the median absolute deviation (MAD) to quantify the location of the main variability component in the line. The wings of the lines seem to be consistent throughout this quiescent episode (although H$\beta$ appears broader), which justifies the use of the wings to trace the orbital motion of the WD; but at low-velocities we observe a narrower component (FWHM$\sim850$ \kms) which drives most of the line variability. This component seems to be filling the core of the line, arising from an additional asymmetry masking the characteristic double-peaked pattern of the accretion disc  \citep[see][]{HorneMarsh:1986}. %This component is very evident on the Doppler tomography presented in Section~\ref{sec:tomo} where no clear signal of the disc is present.

The low-velocity feature is rather unique. A usual culprit -- if it is indeed emitted in the plane of the accretion disc and within the potential well of the WD -- would be situated outside the tidal radius \citep{Paczynski:1977,Whitehurst:1991} as can be observed in the Doppler tomograms in Fig.~\ref{dopmap-hab} discussed in Section~\ref{sec:tomo}. This is rather unusual in a quiescent state of a CV. Such extended material have unstable orbits and will quickly dissipate the angular momentum, accumulating material at the edge of the disc. However, transient features outside of this tidal radius have been observed as material spreads in/out at the onset of outbursts \citep[e.g.,][]{Neustroev:2019}. The suggestion that the outer disc regions have a very low density allowing the gas stream to flow almost freely before seen as an emission component has been proposed before by \citet{Neustroev:2016}, who find that this occurs in several other cataclysmic variables (see references in their paper). This slow velocity component was present during the 2013 campaign (see Fig.~\ref{fig:photo}) but is absent in the study of \citet{Horne:1986}.

\section{Conclusions}
\label{conclusions}

We have made a radial velocity study of the complex Z Cam-type star AH~Her during a rare long minimum. There is only one previous study of this system published by \citet{Horne:1986}. We derived the semi-amplitudes of the radial velocity curves of both components. In particular, we obtained with great accuracy the value of the semi-amplitude of the secondary  $K_R~=~152~\pm~2$ \kms, and consider this result a landmark on which to base the rest of the physical parameters. We stress the fact that the value found by \citet{Horne:1986}, $K_R =158 \pm 8$ \kms, is consistent within the errors with our own. Our $K_W =121 \pm 4$ \kms\ value is also consistent with that of \citet{Horne:1986}: $K_W =126 \pm 4$ \kms. \par
The co-added observations of the secondary star indicate a spectral type of K5. We discussed in detail why we favour the possibility of the donor star being either moderately evolved or under the regime of extreme assumptions, rather than it being a ZAMS star. Nonetheless, using the $M_1- M_2$ diagram, we explore the masses and inclination angles for all of the three sequences. For a K5 ZAMS star we obtained an inclination angle between $43^\circ$ and $44^\circ$, indicating a mass of $M_R=0.74 M_\odot$ for the secondary, and of $M_w=0.93 M_\odot$ for the primary star. For the moderately evolved sequence and for the extreme assumptions one, we relaxed the condition of a spectral type of K5 and look instead at the range of K4--K6. For both of these sequences we observed that the optimal value for the inclination angle was of $i=48^\circ\pm 2^\circ$, which corresponds to masses of ${ M }_{ w }=0.74 \pm 0.06~ M_{\odot}$ and  ${ M }_{ R }=0.59 \pm 0.05~  M_{\odot}$. \par
%Based on the range in spectral type of the secondary and from the $M_1- M_2$ diagram, we derived an inclination angle higher than that previously obtained by \citet{Horne:1986}, and thus the masses we obtain are: $M_{w}=0.77 \pm 0.03 M_{\odot}$ and ${ M }_{ R }=0.57 \pm 0.04  M_{\odot}$, which are smaller than those obtained by these last authors. 
The rotational velocity of the secondary derived from the co-rotation condition has been compared with a method in which we broadened the visible absorption lines. The results coincide within the errors. \par
Doppler tomography of H$\alpha$ and H$\beta$ show that the emission is concentrated in a large asymmetric region at low velocities (in a position opposite to the donor star as found in the SW~Sex stars), outside the tidal radius and therefore at an unstable position. We also compare the H$\alpha$ and H$\beta$ profile lines with the work by \citet{Horne:1986} and show that although the wings are very similar, the core of the line in our case is single-peaked instead of double-peaked, a difference we attribute to its transient nature.

\section*{Acknowledgements}
The authors acknowledges financial support prom  PAPIIT project IN114917. JER acknowledges support from a ``Leids Kerkhoven-Bosscha Fonds'' (LKBF) travel grant to visit the API at UvA. LJS and ARM would like to thank DGAPA/UNAM for financial support provided by PAPIIT projects IN102517 and IN102617. JVHS acknowledges support from a STFC grant ST/R000824/1. We acknowledge the variable star observations from the AAVSO International Database contributed by observers worldwide and used in this research. This research made use of {\sc astropy}, a community-developed core {\sc python} package for Astronomy \citep{Astropy-Collaboration:2013aa} and {\sc matplotlib} \citep{Hunter:2007aa}. We would like to thank the anonymous referee who has been instrumental in greatly improving this paper.

\section*{Data availability}
The data underlying this article will be shared upon reasonable request to the corresponding author.



\bibliographystyle{mnras}
\bibliography{bibliography.bib} % if your bibtex file is called example.bib

%%%%%%%%%%%%%%%%% APPENDICES %%%%%%%%%%%%%%%%%%%%%

% Don't change these lines
\bsp	% typesetting comment
\label{lastpage}
\end{document}